\documentclass[10pt,letterpaper]{article}
\usepackage[top=0.85in,left=2.75in,footskip=0.75in]{geometry}

\usepackage{amsmath,amssymb}

\usepackage{changepage}

\usepackage[utf8x]{inputenc}

\usepackage{textcomp,marvosym}

\usepackage{cite}

\usepackage{nameref,hyperref}

\usepackage[right]{lineno}

\usepackage{microtype}
\DisableLigatures[f]{encoding = *, family = * }

\usepackage[table]{xcolor}

\usepackage{array}

\newcolumntype{+}{!{\vrule width 2pt}}

\newlength\savedwidth

\usepackage{setspace} 
\raggedright
\usepackage[aboveskip=1pt,labelfont=bf,labelsep=period,justification=raggedright,singlelinecheck=off]{caption}

\makeatletter
\renewcommand{\@biblabel}[1]{\quad#1.}
\makeatother

\usepackage{verbatim,shellesc}

\usepackage{lastpage,fancyhdr,graphicx}
\usepackage{epstopdf}
\fancyheadoffset[L]{2.25in}
\fancyfootoffset[L]{2.25in}
\begin{document}
\vspace*{0.2in}

\begin{flushleft}
{\Large
\textbf\newline{Bots, disinformation, and the first impeachment of U.S. President Donald Trump} 
}
\newline
\\
Michael Rossetti\textsuperscript{1,2\Yinyang * },
Tauhid Zaman\textsuperscript{3\Yinyang}
\\
\bigskip
\bigskip
\textbf{1} Operations and Information Management, Georgetown University, Washington, District of Columbia, United States of America
\\
\textbf{2} Technology, Operations, and Statistics, New York University, New York, New York, United States of America
\\
\textbf{3} Operations Research, Yale University, New Haven, Connecticut, United States of America
\\
\bigskip
\bigskip

\Yinyang These authors contributed equally to this work.
\\
\bigskip
\bigskip

* Corresponding author 

Email: mjr300@georgetown.edu (MR)

\end{flushleft}

\newpage


\section*{Abstract}

Automated social media accounts, known as bots, have been shown to spread disinformation and manipulate online discussions.  We study the behavior of retweet bots on Twitter during the first impeachment of U.S. President Donald Trump. We collect over 67.7 million impeachment related tweets from 3.6 million users, along with their 53.6 million edge follower network. We find although bots represent 1\% of all users, they generate over 31\% of all impeachment related tweets. We also find bots share more disinformation, but use less toxic language than other users. Among supporters of the Qanon conspiracy theory, a popular disinformation campaign, bots have a prevalence near 10\%. The follower network of Qanon supporters exhibits a hierarchical structure, with bots acting as central hubs surrounded by isolated humans. We quantify bot impact using the generalized harmonic influence centrality measure.  We find there are a greater number of pro-Trump bots, but on a per bot basis, anti-Trump and pro-Trump bots have similar impact, while Qanon bots have less impact. This lower impact is due to the homophily of the Qanon follower network, suggesting this disinformation is spread mostly within online echo-chambers.



\section*{Introduction}


On December 18, 2019, the United States House of Representatives voted to approve articles of impeachment against President Donald Trump.  The resulting trial in the Senate concluded on February 5, 2020 when the Senate voted to acquit the president. During this period, online social media platforms became a battlefield for information warfare between supporters and opponents of the president \cite{roose2019brace}. 


While much of the activity originated from users engaging in genuine political debate, a significant proportion came from accounts known as \emph{bots}. 
Social bots are automated social media accounts programmed to share certain content and interact with other users \cite{ferrara2016rise}.  
Different kinds of bots are programmed for different purposes, including
 traditional spam bots which aggregate news content or distribute links \cite{ferrara2016rise},
 financial bots which advertise commercial products or attempt to influence financial markets \cite{ferrara2016rise}, ``astroturf'' bots which promote political figures and their policies \cite{ferrara2016rise, sayyadiharikandeh2020detection},
and fake follower accounts used for inflating a user's follower network \cite{ferrara2016rise}. 



Due to their automation, social bots have the potential to manipulate social media discussions \cite{shao2018spread,ferrara2016rise,varol2017bots}. 
The use of bots to spread targeted political messages online, known as computational propaganda \cite{woolley2018computational}, has been growing in recent years.  
Studies have chronicled efforts by bots to manipulate online discussions surrounding U.S. elections and political events since 2016 \cite{russianbots,shane2017fake,guilbeault2016twitter,byrnes2016bot, ferrara2017disinformation, bessi2016social,ferrara2020characterizing}.  

Recent trends show bots amplifying and spreading false or misleading news stories and conspiracy theories (collectively known as \textit{disinformation}).
The phenomenon of bots spreading disinformation has been observed in online discussions ranging from U.S. elections \cite{ferrara2020characterizing} to public health issues such as vaccines \cite{walter2020russian,broniatowski2018weaponized} and the COVID-19 pandemic \cite{ferrara2020types}. 
The introduction of disinformation makes the risk posed by bots much greater, as it allows bots to create false narratives that take hold with a large population, resulting in dangerous outcomes. 
For example: on December 4, 2016, an armed believer of the ``Pizzagate'' conspiracy theory fired his weapon in a {D.C. pizza parlor} \cite{zuckerman2019qanon, pizzagatelegacy, pizzagateanatomy, pizzagateattack}; 
and on January 6, 2021 a mob of Trump supporters influenced by the ``Qanon'' conspiracy theory would storm the U.S. Capitol building in an effort to forcibly overturn the results of the 2020 U.S. presidential election \cite{roose2021qanon, tollefson2021qanon, xu2022network}.
Due to the unique threat bots pose, it is important to be able to identify the bots and quantify their impact in spreading disinformation. 



There is a large body of research on bot detection.  Many bot detection methods rely on analyzing characteristics and behaviors of individual accounts.  Some methods in this category focus on the temporal behavior of accounts \cite{ferraz2015rsc, zhang2011detecting }.  Others focus on the text posted by the accounts \cite{kudugunta2018deep, igawa2016account, clark2016sifting}.  Many approaches combine a variety of features of an account and use them as input to a machine learning classifier \cite{morstatter2016new, gilani2017classification, yang2020scalable, dickerson2014using}.
The most prominent method in this class  is the Botometer (formerly BotOrNot) \cite{davis2016botornot} which uses a machine learning model applied to all data of an account (tweets, profile, followers).  Botometer has been used in a variety of academic studies of social media bots \cite{bessi2016social, ferrara2017disinformation, broniatowski2018weaponized, shao2018spread, luceri2019red, pozzana2020measuring  }.

It has been observed that methods which look at accounts individually may not be able to detect evidence of coordination between groups of accounts \cite{cresci2017paradigm, mazza2019rtbust, yang2020scalable}.
For this reason, other methods have been developed which focus on analyzing collective behaviors and similarities among groups of accounts \cite{cresci2017paradigm, cresci2016dna, vo2017revealing, mazza2019rtbust, des2021detecting}, or including such analysis in a hybrid ``tiered'' approach \cite{beskow2019its}.  
It has been found that some of these collective behavior based methods outperform methods which look at accounts individually \cite{des2021detecting}.


In terms of attempts to quantify bot impact, existing studies provide simple statistics such as the number of bots or volume of content they share \cite{bessi2016social, ferrara2017disinformation}.
However, these statistics do not incorporate the interaction of  social network structure, user activity levels, and user sentiment.  
Some studies have looked individually at the positioning of bots within the social network \cite{shao2018anatomy} or the bot retweet response time \cite{shao2018spread}, but not the interaction of these factors.  
Recent work has presented a novel network centrality measure known as \emph{generalized harmonic influence centrality} that combines all of these factors to assess bot impact in online political discussions \cite{des2021detecting, hunter2018opinion}. 
This centrality measure provides a better measure of bot impact than individually examining each factor. 


In this study, we focus on social bots discussing the first impeachment of U.S. President Donald Trump on Twitter. We find that bots are 66 times more active than normal human users, producing nearly one third of all impeachment-related content, despite representing less than 1\% of all accounts discussing the impeachment.  Bots tend to share news from lower quality sources (including disinformation) than their co-partisan human counterparts.  Bots have an unusually high prevalence among Qanon conspiracy supporters, indicating that efforts are being made to artificially amplify this disinformation.
Using generalized harmonic influence centrality, we show that pro-Trump and anti-Trump bots have a similar level of impact per bot, but Qanon bots have a lower per bot impact.  Analysis of the Qanon follower network suggests this lower impact is due to Qanon users existing in an online echo-chamber with a high amount of ideological homogeneity.


\section*{Results}
\begin{figure}[!h]
\includegraphics[width=\textwidth]{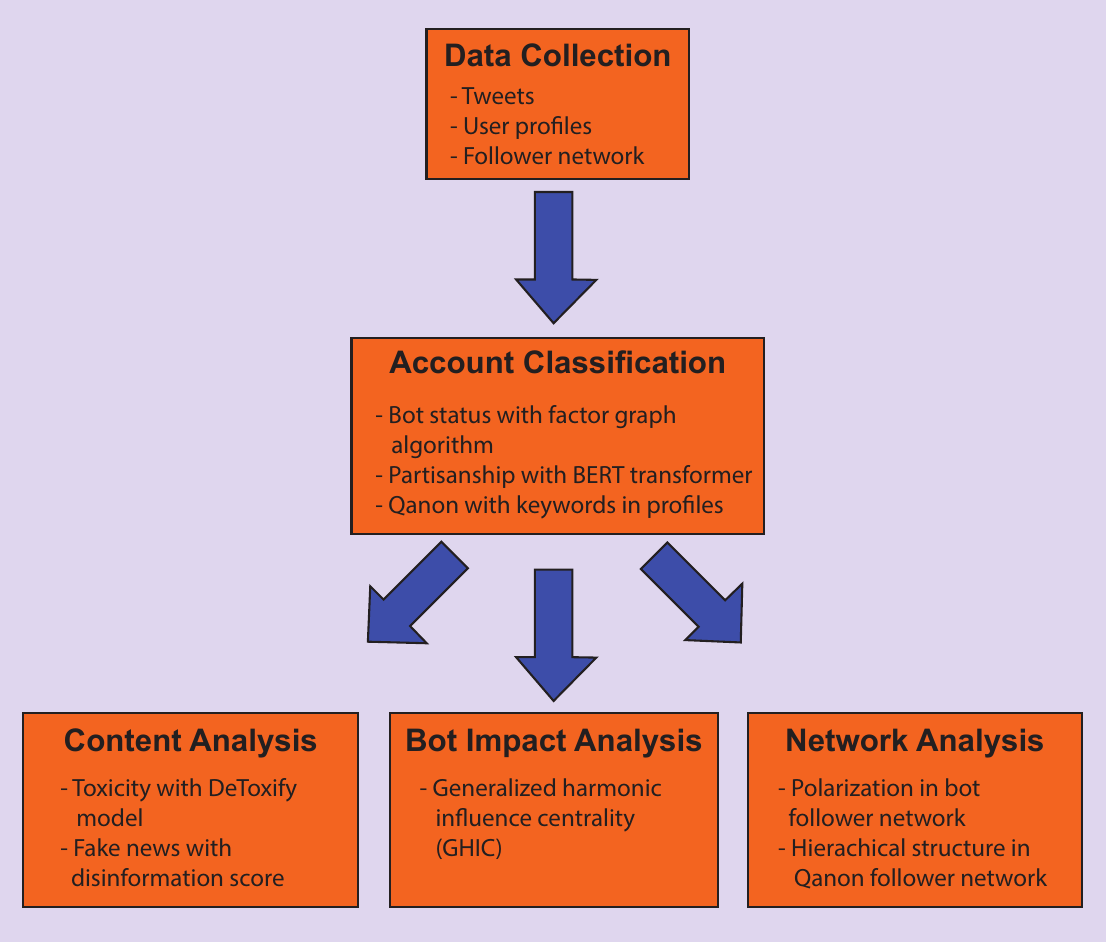}
\caption{{\bf Diagram of the different steps taken in our analysis of bots and disinformation in the Twitter discussion surrounding Donald Trump's first impeachment.}}
\label{fig:pipeline}
\end{figure}

Our analysis contains multiple steps, which are shown in Fig~\ref{fig:pipeline}.  We begin by collecting social media data related to the impeachment.  Then we apply a variety of methods to classify the accounts by partisanship, bot status, and Qanon support.  We perform a variety of analyses comparing the different types of accounts in terms of their posted content and network structure.  Finally, we quantify the impact of the bots using generalized harmonic influence centrality.

\subsection*{Data collection}
We constructed a list of keywords and hashtags related to the first impeachment of President Donald Trump.  This list included partisan terms advocating for each side in the discussion, and non-partisan terms related to developments in the impeachment  (see Methods). From  December 12, 2019 to March 24, 2020, we collected tweets which contained at least one of these keywords.  This collection process  resulted in a dataset of 67.7 million tweets posted by 3.6 million unique Twitter accounts.  We also collected the user profiles of these accounts and their 53.6 million edge follower network (see Methods).  We provide summary statistics of this dataset in Table~\ref{table1}.

\subsection*{Account classification}
We classified the Twitter accounts into different groups which characterized different aspects of their preferences and behavior.  These included political partisanship, support for the Qanon movement, and whether or not the account was an automated bot.  Each group label was assigned using a different method.  A summary of the group statistics is provided in Table~\ref{table1}.

\begin{table}[h!]
    \begin{center}
  \caption{ \bf Data collection results, by account type.} 
  \label{table1}
    \begin{tabular}{|c|c|c|r|r|}
  \hline
  \textbf{Partisanship} & \textbf{Bot status} & \textbf{Qanon status} & \textbf{Accounts} & \textbf{Tweets}\\\hline
  Anti-Trump & Human & Normal & 2,273,831 & 25,103,340\\\hline
  Pro-Trump	&Human & Normal & 1,279,638& 18,717,915	\\\hline
Pro-Trump &	Human &	Q-anon & 22,926 & 2,845,521 \\\hline
Pro-Trump &	 Bot & Normal & 11,571 & 9,880,481\\\hline
Anti-Trump & Bot &	Normal &	10,145 &	9,220,258\\\hline
Pro-Trump &	Bot	& Q-anon &	2,434 &	1,899,042 \\\hline
       &&& \textbf{3,600,545} &	\textbf{67,666,557}\\\hline
  \end{tabular}
    \end{center}
\begin{flushleft} Statistics of account types and tweets in Twitter dataset for the first impeachment of U.S. President Donald Trump.
\end{flushleft}
\end{table}

First, we considered the political partisanship of the accounts.  We  trained a bidirectional encoder representations from transformers (BERT) model to measure the political partisanship of any given tweet text \cite{devlin2018bert}.  The model was trained using a subset of the impeachment tweets for which we were able to assign ground-truth labels (see Methods).  
A label of zero represents strong anti-Trump sentiment and a label of one represents strong pro-Trump sentiment.  We used the trained model to measure the sentiment of all tweets in our dataset.  Then, we assigned a partisanship score to each Twitter account equal to the mean value of the partisanship score of their tweets.  The accounts were labeled  anti- (pro-) Trump if their partisanship score was less than or equal to (greater than) 0.5. In total, our dataset had 2.3 million anti-Trump accounts and  1.3 million pro-Trump accounts. This left leaning bias in the Twitter conversation aligns with findings from previous studies \cite{des2021detecting}.

We next identified accounts who were supporters of the Qanon movement, based on those who promoted Qanon hashtags used in their profile descriptions (see Methods).
Using this process, we identified 25,360 Qanon accounts.

To identify bots we used an algorithm based on a factor graph model \cite{des2021detecting}.  This algorithm simultaneously detects multiple bots among a set of Twitter accounts, using only their retweet network (see Methods).  We chose this algorithm because it had minimal data requirements compared to other algorithms \cite{davis2016botornot} (as we are easily able to construct the retweet network from the collected tweets), and also because it exhibited performance similar to or better than other algorithms \cite{des2021detecting}.  We applied the bot detection algorithm to daily retweet networks constructed from tweets posted each day.  This allowed us to identify daily sets of active bots. The algorithm we use is designed to identify bots that engage in excessive retweeting.  These retweet bots produce a disproportionate volume of social media content that can distort online discussions, so we focus our analysis on this type of bot.  Our detection algorithm will not identify bots which exhibit different behavioral patterns that do not consist of excessive retweeting.  Therefore, when we classify an account as a bot, we specifically mean a bot retweeting at an unusually high rate. We note that one limitation of our bot detection algorithm is that it relies upon the retweet network, and so can only detect bots who retweet someone.  However, since our focus is on retweet bots that amplify certain voices, this limitation is not a major issue. 

Our final set of bots consisted of the union of these daily bot sets.  In total we found 24,150 bots, of which 10,145 were anti-Trump and 14,005 were pro-Trump.
This ratio of bots to humans aligns with other studies of U.S. politics on Twitter \cite{ferrara2020characterizing,des2021detecting}. 

\begin{figure}[!h]
\includegraphics[width = \textwidth]{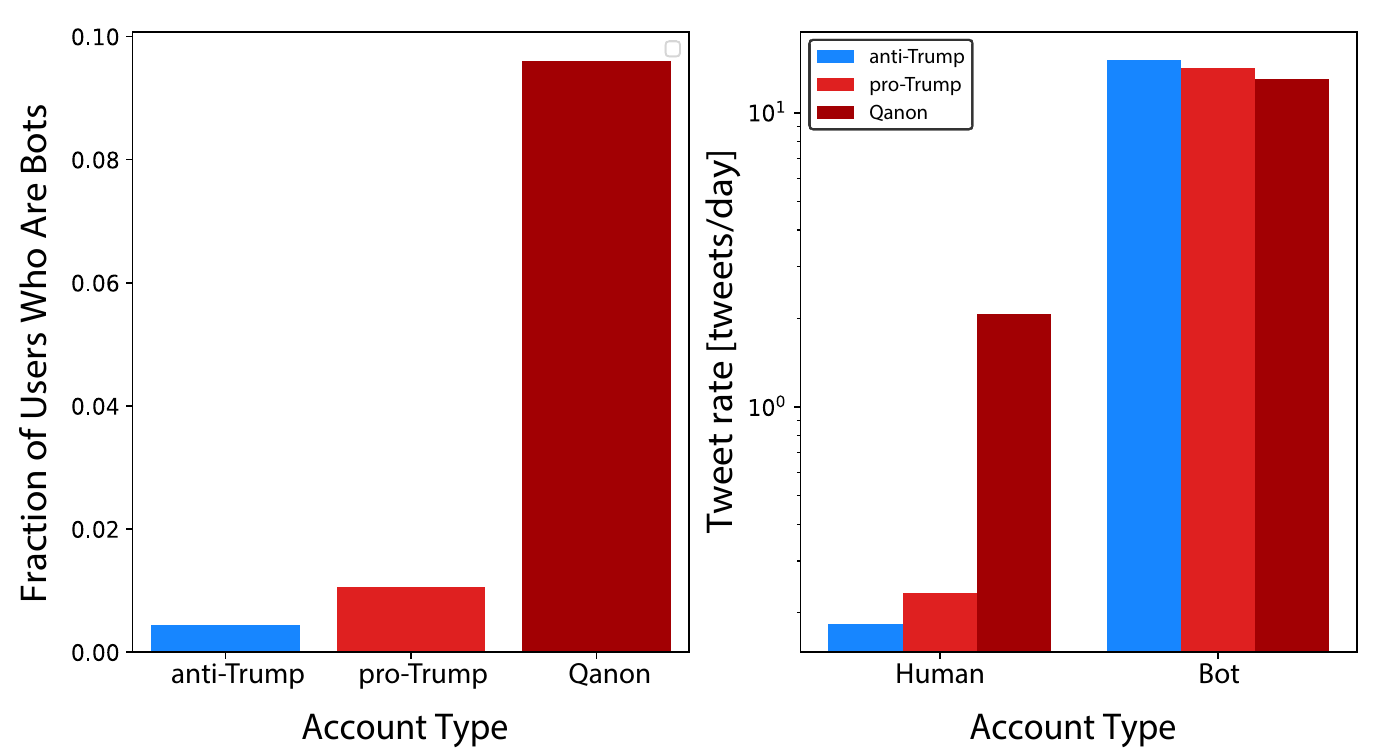}
\caption{{\bf Bot prevalence and activity rates.}
(left) Fraction of  bots and (right) average tweet rate of each account type category.}
\label{fig1}
\end{figure}

We summarize the prevalence of bots in different groups in Fig~\ref{fig1}.  We see that bots have a slightly higher prevalence among pro-Trump accounts than anti-Trump accounts (p-value $<10^{-6}$).  However, the bot prevalence among Qanon supporters is nearly an order of magnitude larger than it is among normal accounts (p-value $<10^{-6}$).  
This suggests malicious actors may be attempting to use artificial accounts to amplify Qanon content.  
On one hand, this may be reassuring, as it suggests 
there aren't as many real Qanon supporters on social media as there may appear.
On the other hand, significant bot presence means that Qanon content is being spread at a higher rate and with a potentially higher reach than could be achieved with humans alone.  
This high bot fraction is even more concerning when looking at the tweet rate of the accounts in Fig~\ref{fig1}, as  Qanon human supporters tweet approximately ten times more frequently than regular humans (p-value $<10^{-6}$), and bots tweet approximately one hundred times more frequently than regular humans (p-value $<10^{-6}$).  
Having a large number of Qanon bots can lead to an unusually large amount of Qanon content being spread through Twitter.  This potentially enhanced reach increases the risk posed by an already dangerous ideology.

\subsection*{Content analysis} 
There are differences in the nature of content posted by the various account types.  We first considered the quality of news stories shared by the accounts.   We leveraged a previously published set of 60 news sites (20 mainstream, 20 hyper-partisan, and 20 fake news, with liberal and conservative leaning sites in each category) whose trustworthiness had been rated by eight professional fact-checkers \cite{pennycook2019fighting}.
We followed the approach used in prior work \cite{pennycook2021shifting, mosleh2021perverse} and calculated a media quality score for each user by averaging the trustworthiness ratings of any of their impeachment related tweets that contained links to any of those 60 sites. There was a link of some sort in  11\% of the tweets, and of those with links, 12\% had a link from one of the 60 news sites.  In total, we were able to calculate a media quality score for  217,692 users.   The media quality score ranges from one to five, with higher values indicating higher trustworthiness.  Like other researchers in this space \cite{guess2019less, grinberg2019fake, pennycook2021shifting, mosleh2021perverse}, we use source trustworthiness as a proxy for article accuracy, because it is not as feasible to rate the accuracy of every shared link.  We note that our analysis of news quality is based on accounts in our dataset who share news stories, so care must be taken when generalizing to a broader set of accounts.  Our conclusions apply only to the subset of accounts of each type who choose to share news stories.  

\begin{figure}[!h]
\includegraphics[width = \textwidth]{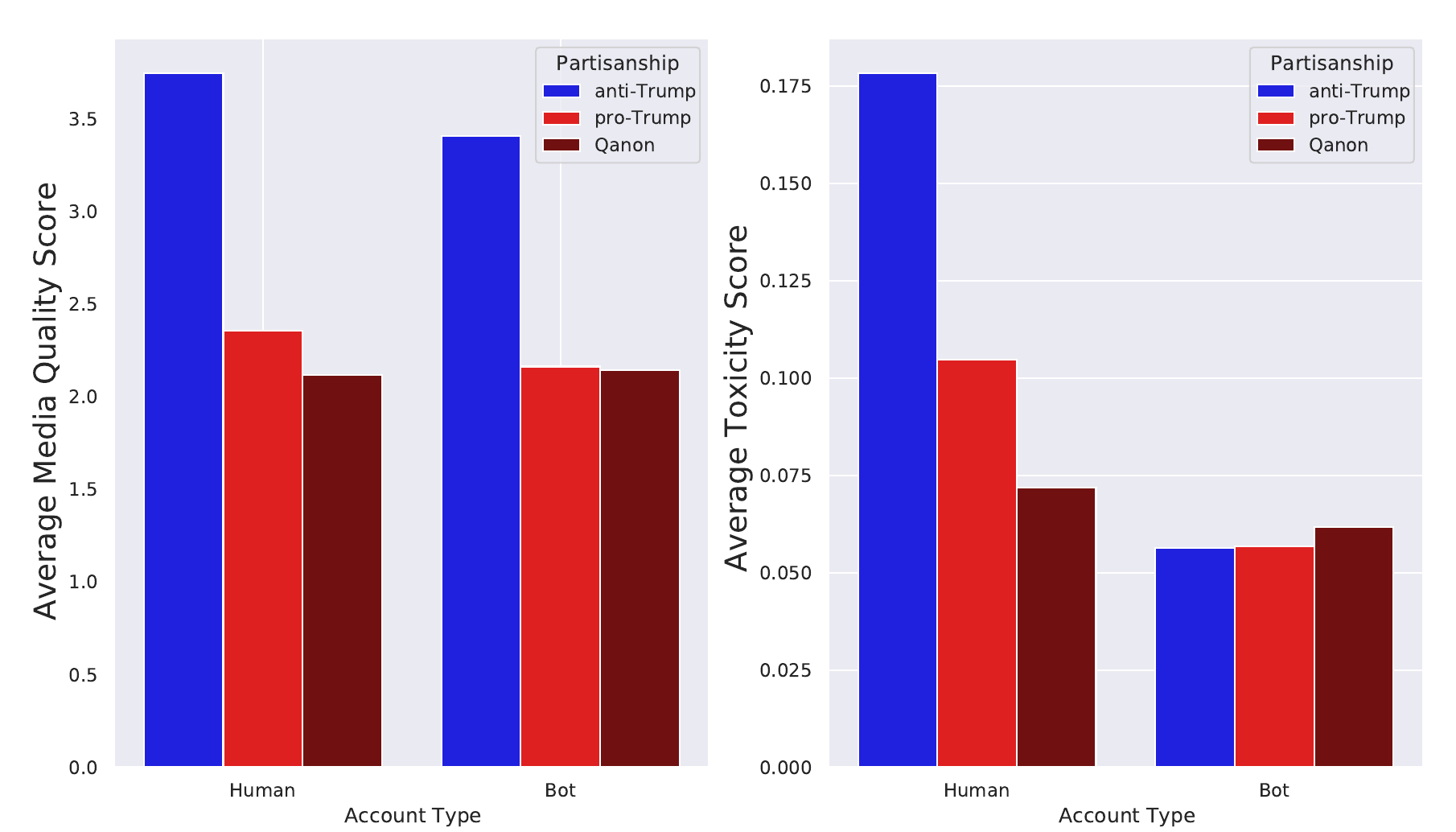}
\caption{{\bf Bot language and content analysis.}
(left)  Average media quality score and (right) average toxicity score of tweets posted by different account types.}
\label{fig2}
\end{figure}

Fig~\ref{fig2} shows the average media quality score of the different account types. The first striking observation is that the media quality score is much higher for anti-Trump accounts than pro-Trump accounts.  This is consistent with past work which found that pro-Trump (Republican) users  were much more likely to share news from untrustworthy news sites than anti-Trump (Democratic) users \cite{grinberg2019fake}.    We also find that within each partisan group, the media quality score of bots is lower than that of humans {(p-value $<10^{-6}$ for bots versus humans in each partisan group)}.  Qanon humans have a lower media quality score than both normal pro-Trump humans {(p-value $<10^{-6}$)} and normal pro-Trump bots (p-value = 0.0002).  However, we find no statistical difference between the average media quality score for Qanon humans and Qanon bots (p-value = 0.29).

In addition to what news media the accounts are sharing, we also investigate the tone and sentiment of the tweets they post.  One measure of this is known as toxicity, which captures how harmful or unpleasant a tweet is.  
We measured the toxicity of all tweets in our dataset using Detoxify \cite{hanu2021detoxify}, a neural network model for toxicity detection that has achieved similar performance to top scorers in multiple Kaggle competitions related to detecting toxic language in web-based content \cite{github_detoxify}. Detoxify has been used in several studies for profiling hate speech spreaders \cite{huertas2021profiling}, detecting cyberbullying in online forums \cite{vo2021automatically},  establishing toxicity thresholds for content \cite{iqbal2022exploring}, manipulating language models \cite{bagdasaryan2022spinning}, video-text retrieval systems \cite{hanu2022vtc},  and measuring the safety of conversational neural network models \cite{sun2021safety}. 

Despite its high performance, the model is sensitive to insults and profanity, which may result in higher toxicity scores regardless of the intent or tone of the message.
The model produces toxicity scores ranging from zero to one, with higher values indicating higher toxicity.  We applied Detoxify directly to the tweet text.  If the tweet is a retweet, then Detoxify is applied to the text of the retweet.  We do not open any links in a tweet or measure the toxicity of the destination website.  We feel this is sufficient as a link does not have the same impact on an online conversation as toxic text in a tweet.

The average toxicity score for each account type is shown in Fig~\ref{fig2}.  We note that the distribution of toxicity scores for each group is characterized by two clusters located near zero and one.  The cluster near one contains the toxic users which set the average toxicity of the group. Further analysis of the distribution of these scores is provided in the Materials and Methods. One striking observation in Fig~\ref{fig2} is that anti-Trump humans have the highest average toxicity score by a wide margin.  The next highest toxicity group are pro-Trump humans. Qanon humans and bots have lower toxicity than normal humans {(p-value $<10^{-6}$)}. Bots of each partisanship group have lower toxicity levels than their co-partisan human counterparts {(p-value $<10^{-6}$)}. The high toxicity levels of anti-Trump humans may be due to their outrage about Trump's actions, and their disagreement with his acquittal.  
Conversely, low toxicity levels among pro-Trump users may reflect their attempts to post positive content in a show of support for the president.

When we focus on the bot accounts, we find two general patterns.  Within each partisan group, bots share lower quality media than humans and post less toxic content than humans.  From this we can deduce that bots mainly share low quality news stories (relative to their human counterparts within their partisan group), but they {tend} not to amplify negative messages which use aggressive language.  This suggests that the bots are more focused on spreading information and not on agitating users with toxic posts.

Finally, we study 
bot retweet patterns.
A retweet occurs when one account re-posts the content of another, thus sharing the original content with all of their followers.
High retweet counts indicate high levels of popularity for the original tweet's content and author. 
Tables ~\ref{table_statuses_most_retweeted_0} and ~\ref{table_statuses_most_retweeted_1} show the content most retweeted by anti- and pro-Trump bot accounts, respectively, and the number of bots who retweeted each message. These top retweeted messages are primarily authored by elected officials involved in the impeachment proceedings.

\begin{table}[h!]
	\begin{center}
		\caption{ \bf Statuses most retweeted by anti-Trump bots.}
		\label{table_statuses_most_retweeted_0}
		\begin{tabular}{|p{0.7\textwidth}|p{0.2\textwidth}|}
		\hline
  
		\centering\textbf{Retweeted status text} & \textbf{Bot count}\\\hline

                \textit{RT \@@SpeakerPelosi: The House cannot choose our impeachment managers until we know what sort of trial the Senate will conduct. President Trump blocked his own witnesses and documents from the House, and from the American people, on phony complaints about the House process. What is his excuse now?} & 4,983 \\\hline 
                \textit{RT \@@RepAdamSchiff: Lt. Col. Vindman did his job.  As a soldier in Iraq, he received a Purple Heart. Then he displayed another rare form of bravery — moral courage. He complied with a subpoena and told the truth. He upheld his oath when others would not. Right matters to him. And to us.} & 4,225  \\\hline 
                \textit{RT \@@RepAdamSchiff: Impeachment of a president is a serious undertaking. The Senate’s role is to act as an impartial jury and provide a fair trial. Fair to the President and to the American people. That means seeing all the evidence, documents and witnesses. What is McConnell afraid of?} & 4,212  \\\hline 
                \textit{RT \@@SpeakerPelosi: In the Clinton impeachment process, 66 witnesses were allowed to testify including 3 in the Senate trial, and 90,000 pages of documents were turned over. Trump was too afraid to let any of his top aides testify \& covered up every single document. The Senate must \#EndTheCoverUp} & 3,977  \\\hline 
                \textit{RT \@@SpeakerPelosi: The President \& Sen. McConnell have run out of excuses. They must allow key witnesses to testify, and produce the documents Trump has blocked, so Americans can see the facts for themselves. The Senate cannot be complicit in the President's cover-up. \#DefendOurDemocracy} & 3,955  \\\hline

	\end{tabular}
	\end{center}
    \begin{flushleft} The messages retweeted by the greatest number of anti-Trump bots, and the number of anti-Trump bots who retweeted each.
    \end{flushleft}
\end{table}

\begin{table}[h!]
	\begin{center}
		\caption{ \bf Statuses most retweeted by pro-Trump bots.}
		\label{table_statuses_most_retweeted_1}
		\begin{tabular}{|p{0.7\textwidth}|p{0.2\textwidth}|}
		\hline
  
		\centering\textbf{Retweeted status text} & \textbf{Bot count}\\\hline

                \textit{RT \@@realDonaldTrump: I was very surprised \& disappointed that Senator Joe Manchin of West Virginia voted against me on the Democrat’s totally partisan Impeachment Hoax. No President has done more for the great people of West Virginia than me (Pensions), and that will....} & 8,265 \\\hline
                \textit{RT \@@realDonaldTrump: ``Nancy Pelosi said, it's not a question of proof, it's a question of allegations! Oh really?'' \@@JudgeJeanine \@@FoxNews What a disgrace this Impeachment Scam is for our great Country!} & 8,134 \\\hline
                \textit{RT \@@realDonaldTrump: As hard as I work, \& as successful as our Country has become with our Economy, our Military \& everything else, it is ashame that the Democrats make us spend so much time \& money on this ridiculous Impeachment Lite Hoax. I should be able to devote all of my time to the REAL USA!} & 8,059 \\\hline
                \textit{RT \@@realDonaldTrump: Crazy Nancy Pelosi should spend more time in her decaying city and less time on the Impeachment Hoax!} & 7,645 \\\hline
                \textit{RT \@@realDonaldTrump: Many believe that by the Senate giving credence to a trial based on the no evidence, no crime, read the transcripts, ``no pressure''  Impeachment Hoax, rather than an outright dismissal, it gives the partisan Democrat Witch Hunt credibility that it otherwise does not have. I agree!} & 7,473 \\\hline

	\end{tabular}
	\end{center}
    \begin{flushleft} The messages retweeted by the greatest number of pro-Trump bots, and the number of pro-Trump bots who retweeted each.
    \end{flushleft}
\end{table}

\begin{figure}[!h]
\includegraphics[width = \textwidth]{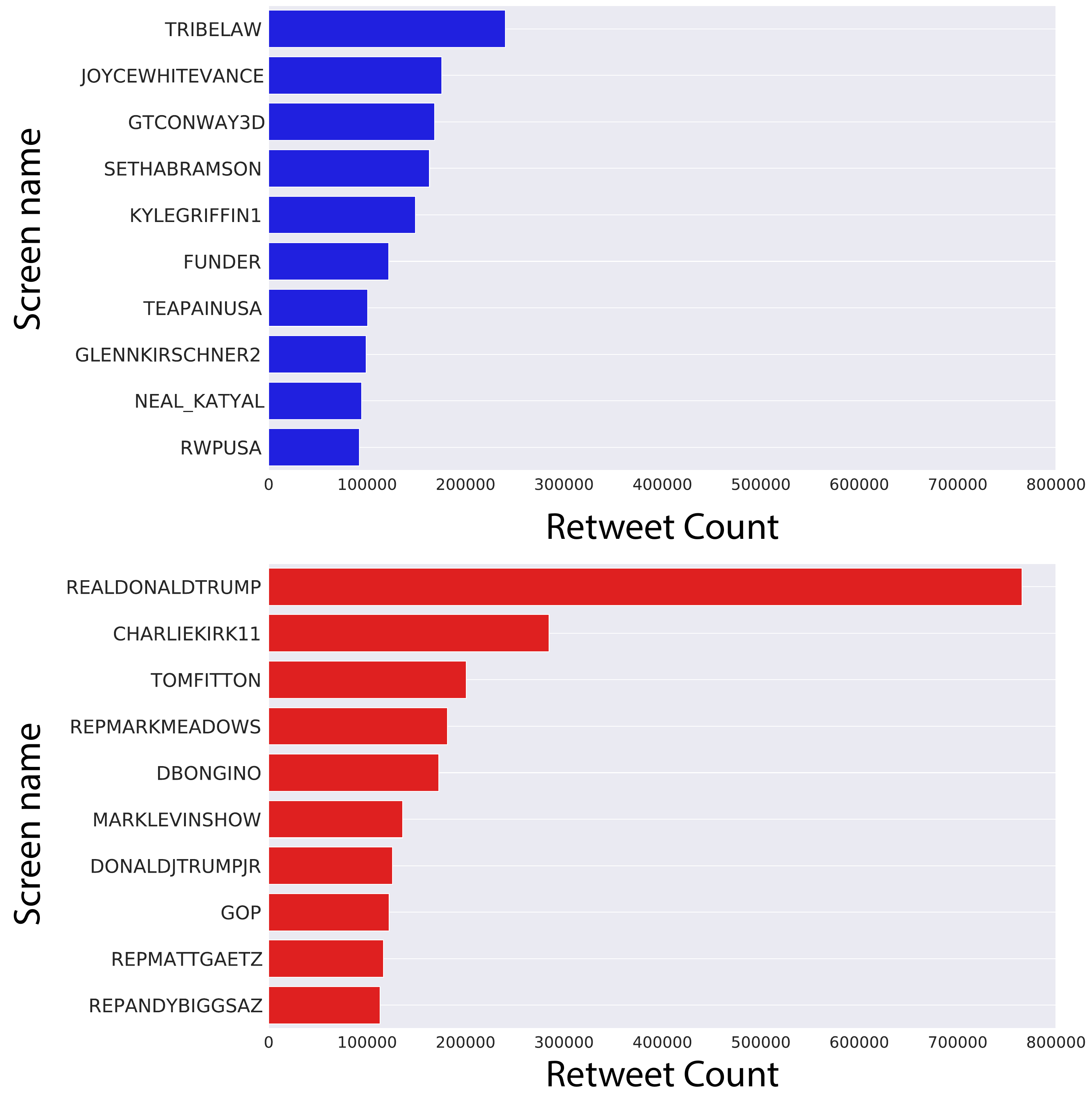}
\caption{{\bf Bot retweet targets.}
Number of retweets by (top) anti-Trump bots and (bottom) pro-Trump bots for top retweeted Twitter accounts.}
\label{fig3}
\end{figure}

Fig~\ref{fig3} shows the accounts most retweeted by anti- and pro-Trump bot accounts, respectively, and the number of retweets each account received from bots.  This allows us to examine which accounts benefited most from bot activity.

Here we see a clear difference in the structure of the bot retweet distributions.  On the anti-Trump side, the bot retweets are distributed in a rather uniform manner over the top ten accounts, with retweet counts ranging from {240,336 to 92,233}.  However, bot retweet counts on the pro-Trump side range from {765,512 to 113,261}, with a very concentrated distribution in which Donald Trump receives the most bot retweets by far, earning more than twice the amount of retweets than the second most retweeted account. 
This suggests Donald Trump is a singular figure in terms of bot retweets.  

It is also interesting to note that the top ten bot retweeted accounts on the anti-Trump side are all political pundits.  There are no elected or government officials.  In contrast, among the top ten accounts on the pro-Trump side, there are three elected officials (President Donald Trump, Congressman Matt Gaetz, Congressman Andy Biggs) and one cabinet official (Chief of Staff Mark Meadows).  Finally, we note that the official Twitter account of the republican party (GOP) is among the top ten accounts retweeted by bots, while the account of the democrat party (DNC) is not.  This analysis suggests that pro-Trump bots are more actively amplifying officials with political or government power than anti-Trump bots.

\subsection*{Network analysis}
\begin{figure}[!h]
\includegraphics[width = \textwidth]{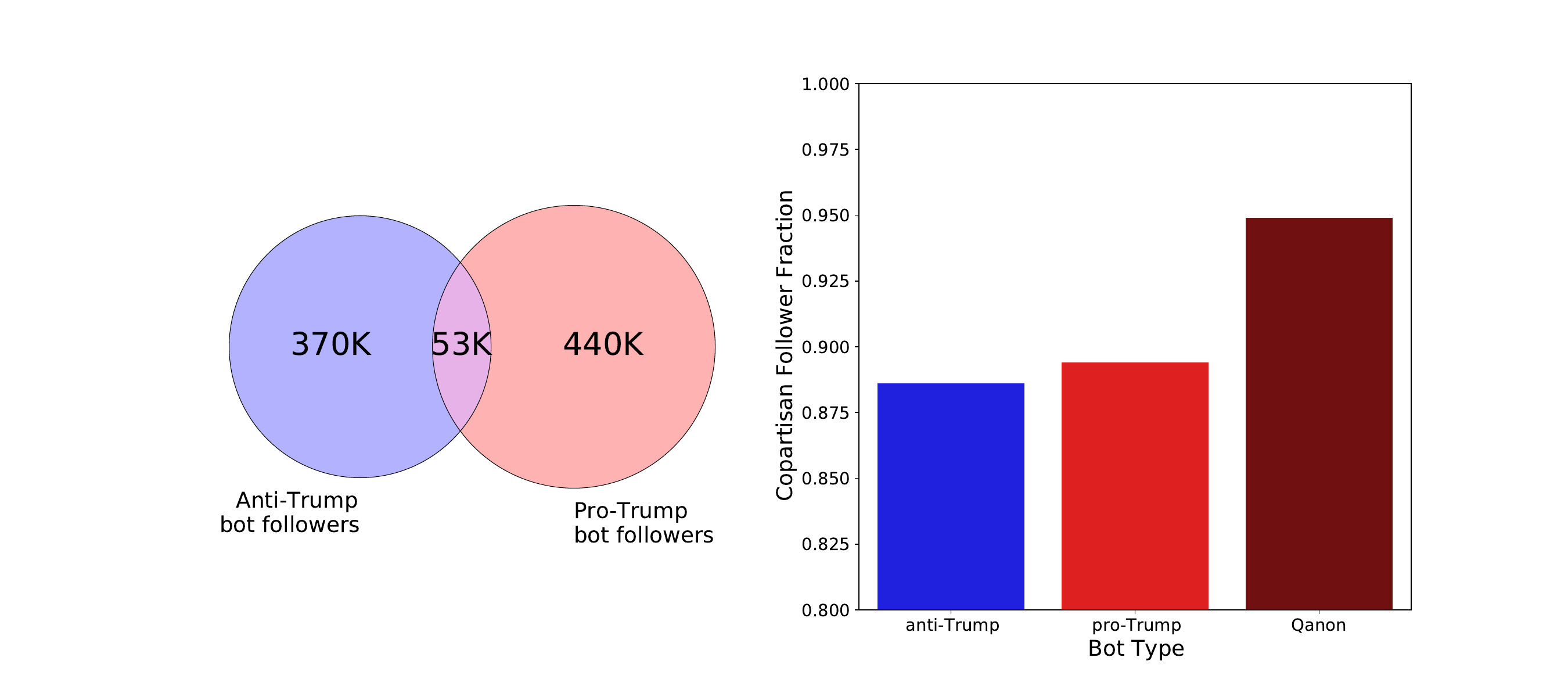}
\caption{{\bf Bot follower network reach and composition.}
(left) Venn diagram of users who follow anti-Trump and pro-Trump bots.  (right) Fraction of co-partisan followers for different bot types.}
\label{fig4}
\end{figure}

\begin{figure}[!h] 
\includegraphics[width = \textwidth]{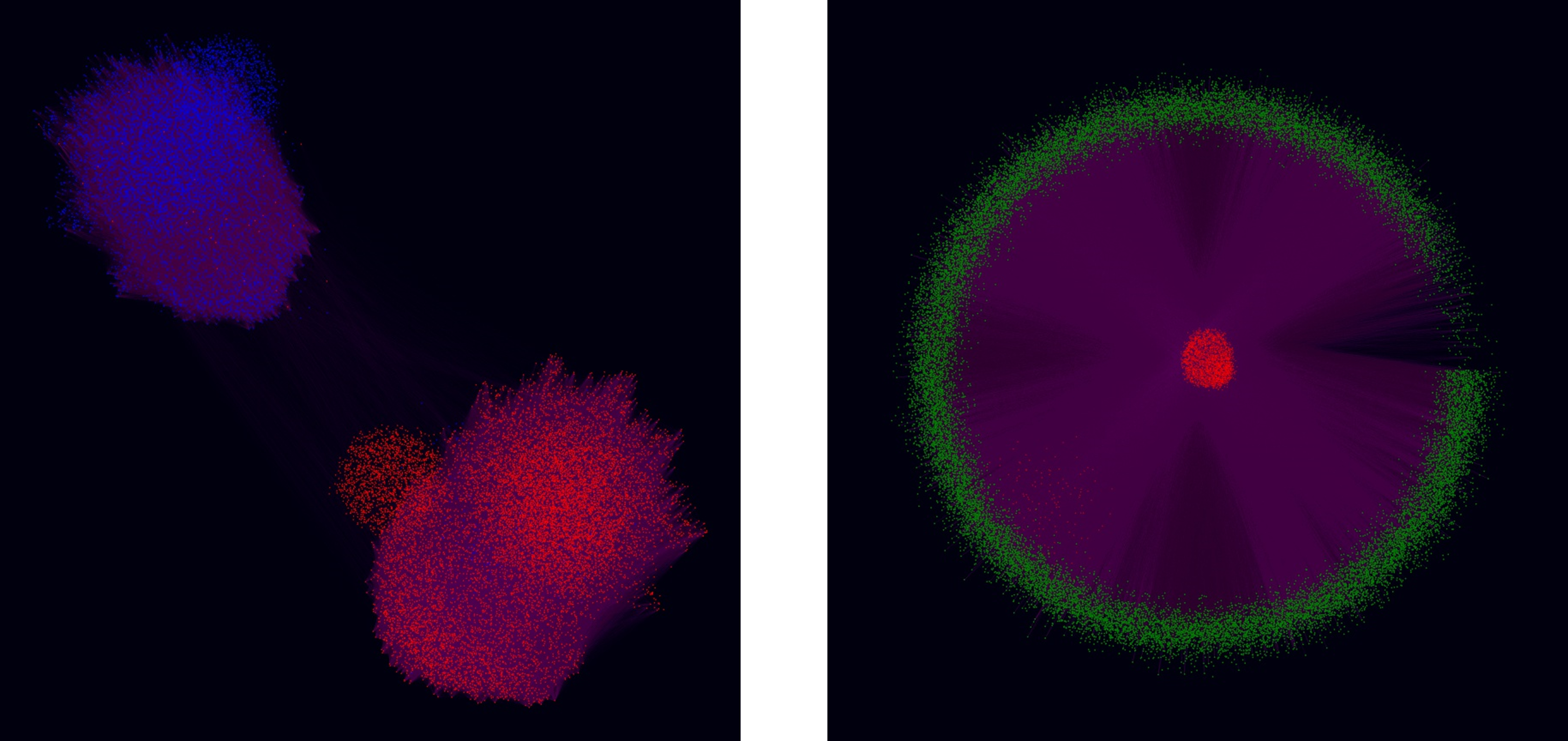}
\caption{{\bf Bot follower network structure.}
(left) Follower network of bot accounts colored by partisanship, where anti-Trump bots are blue, and pro-Trump bots are red.  (right) Follower network of Qanon accounts, where bots are colored red and humans are colored green.}
\label{fig5}
\end{figure}

Bots and Qanon supporters have networks that exhibit distinct properties.  We first consider the bot follower network.  Fig~\ref{fig4} shows the number of unique accounts who follow at least one bot, and the partisanship of the bots they follow.  We find that pro-Trump bots have more followers than anti-Trump bots, which may be due to the pro-Trump bots' greater numbers.  Less than 6\% of the bot followers follow bots in both partisan groups, suggesting that the bot followers' network is highly polarized.  This polarization is very visible in the bots' follower network, as shown in Fig~\ref{fig5}.  Here we see that the two partisan groups of bots are almost totally disconnected.  To obtain a more quantitative measure of this polarization, we calculate the fraction of followers of each bot type who are co-partisan (i.e. those who share the same political affiliation).  A higher value for this measure indicates greater ideological homogeneity in the followers of the bots. Fig~\ref{fig4} shows the co-partisan fractions for each bot type.  We find the values are quite high, being above 0.87 for all bot types.  There is a 1\% difference in the co-partisan follower fraction between the anti-Trump and non-Qanon pro-Trump bots {(p-value $<10^{-6}$)}.  However, the Qanon bots have have a co-partisan fraction that is 6\% to 7\% greater than the non-Qanon bots  {(p-value $<10^{-6}$)}.  This suggests that Qanon bots have an audience that is even more partisan than a standard bot.  

The Qanon follower network, comprised of Qanon bots and humans, has an interesting structure.  We find that this network contains a core of bots which are connected to each other, surrounded by a periphery of humans. Interestingly, these humans are only connected to the bots, and not to each other.  We show this network in Fig~\ref{fig5}, where the nodes are laid out with bots  in the center to highlight this structural property. The connectivity among bots and lack of connectivity among humans suggests there is a hierarchical structure within the Qanon community.  The bots act as sources of content for the humans.  The humans appear not to engage with each other, but rather mainly consume content from the bots.  We will see later that this network structure has implications for the impact of Qanon bots.

\subsection*{Bot impact analysis}
Recall that our focus is on bots that retweet excessively.
We next provide a quantitative measure of the impact of these retweet bots on the impeachment discussion.  Thus far we have presented a detailed analysis of the activity, sentiment, and network structure of the bots.  Impact is a combination of all of these factors.  We would expect more active bots to have more impact, as they post more content.  Bots with larger network reach should also have greater impact.  Finally, bots whose followers are not strong co-partisans would have more impact because they can persuade these followers to their side. In contrast, strong co-partisan followers who already agree with the bots likely cannot be persuaded further.  
A measure that combines these factors is known as harmonic influence centrality \cite{vassio2014message}.  This is a network centrality based on a classic model for opinion dynamics \cite{degroot1974reaching} which incorporates stubborn users with immutable opinions \cite{mobilia2003does}.  Harmonic influence centrality measures how much a set of nodes shifts the average equilibrium opinion in the network. 
The centrality naturally incorporates activity, network reach, and opinions to measure the impact of nodes in a network because it is based on an opinion dynamics model which utilizes these factors.  In its original form, harmonic influence centrality used only the network structure and activity level.  A more recent version, known as generalized harmonic influence centrality (GHIC), incorporates additional data, such as the node sentiment, making it more appropriate for real social networks.  GHIC has been used to quantify the impact of bots in Twitter networks discussing various geo-political events \cite{des2021detecting}.  Because our data is quite similar in nature, we use this generalized version of the measure to quantify the impact of the bots.
 
 To apply GHIC, one first must define the network.  We want to calculate a daily impact measure for the bots to see how their impact evolves over time.  Therefore, the networks we use are daily active follower networks.  For a given day, the daily active follower network is the sub-network of the entire Twitter follower network induced by accounts which are active (post at least one tweet) that day.  We follow the approach in \cite{des2021detecting} to calculate the GHIC of different groups of bots on these networks (see Methods).  
 
\begin{figure}[!h]
\includegraphics[width=\textwidth]{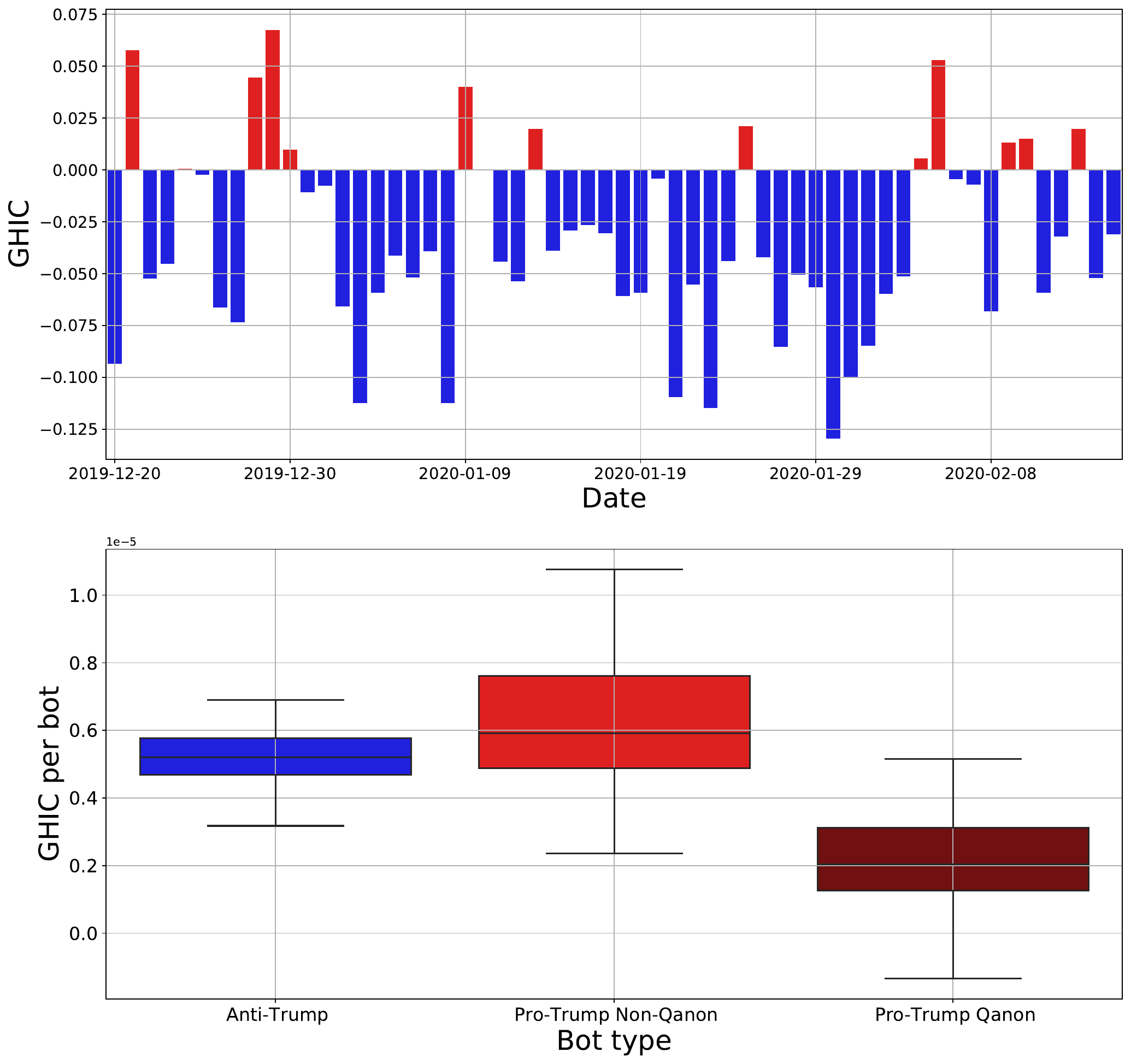}
\caption{{\bf Bot impact analysis.}
(top) Daily generalized harmonic influence centrality (GHIC) score for all bots versus date. (bottom) Boxplot of the daily GHIC per bot for different bot types.}
\label{fig6}
\end{figure}

 We first look at the GHIC of all bots.  The daily GHIC of all bots is shown in Fig~\ref{fig6}, where positive values indicate a shift toward pro-Trump opinions, and negative values indicate a shift toward anti-Trump opinions.  This shows on a given day which partisan side of bots had greater impact. We observe that for most days, the anti-Trump bots have greater impact.    However, there are days when the pro-Trump bots have greater impact.  Upon closer investigation, we find that on these days there was a news story which excited pro-Trump users on Twitter.  For instance, on December 29, 2019 Donald Trump posted a controversial tweet referring to U.S. Speaker of the House of Representatives Nancy Pelosi as ``Crazy Nancy''.  We see that the GHIC had a large positive value that day, indicating that the pro-Trump bots had greater impact. In general, we find that the GHIC achieves a large magnitude on days with polarizing events.  
 
 In addition to the daily GHIC, we are also interested in the efficiency of each group of bots.  By efficiency, we mean the GHIC per bot for a group of bots.  This measure would identify which 
 group of bots are more impactful, on average.
 We consider three groups of bots: anti-Trump, pro-Trump non-Qanon, and pro-Trump Qanon.  For each group, we calculate their daily GHIC per bot for all days.  The boxplot in Fig~\ref{fig6} shows the distribution of the daily GHIC per bot for each group. We see that both groups of non-Qanon bots have very similar GHIC per bot distributions, though the pro-Trump bots have a slightly higher mean value.  However, statistical tests do not provide strong evidence that the means of these two groups are different (p-value =0.05).  In contrast, we find strong evidence that Qanon bots have a lower mean GHIC per bot than non-Qanon bots (p-value $<10^{-6}$).  To understand why Qanon bots are less efficient, we look at Fig~\ref{fig5}.  We saw there that Qanon bots have a higher fraction of co-partisan followers than non-Qanon bots, meaning that Qanon bots are more ideologically aligned with their followers.  Because of this, these bots cannot impact their followers' opinions as much, which  lowers their GHIC efficiency.  Our finding suggests that Qanon bots are not as effective at persuasion.  Rather, they likely preach to a converted audience, which diminishes their impact compared to non-Qanon bots.


\section*{Discussion}

Our study of the Twitter discussion surrounding the first impeachment of Donald Trump found that a small number of bots generated a disproportionately large amount of content.  In addition, the combined follower reach of these bots is extensive.  A primary bot activity is to spread news, and we found that bots generally spread lower quality news than humans.  However, the language used in bot tweets is generally less toxic than that of humans.  Using the GHIC measure we are able to quantify the daily impact of the bots.  We found that bot impact is highest on days with politically charged events.  Overall, anti-Trump bots have a greater impact, but their per bot impact is similar to the pro-Trump bots.  Qanon bots have a lower per bot impact than the other bots.  This is likely due to their high fraction of co-partisan followers, which limits their persuasive ability.

The excessive reach and activity level of bots, combined with their propensity to share news from low quality sources, are cause for concern.  A small number of bots can amplify certain stories or narratives, causing them to reach a large audience. Bots seem to have the greatest impact on days when there is a large amount of partisan agitation, suggesting that bots may be increasing online polarization.   However, one encouraging finding is that the Qanon bots, who spread a particularly dangerous form of disinformation, exist within strong echo-chambers, and as a result have less impact than normal bots.


\section*{Materials and methods}

\subsection*{Data collection}
The  keywords and hashtags used as search criterion for tweets related to the first impeachment of President Donald Trump are shown in Table~\ref{table2}.  
Some of these words were added to the collection list as news stories developed.  
The table shows the date each term was added. 
The Twitter Streaming API 
was used to collect in real time any tweets containing at least one of these terms.
After an initial trial run from December 12 to 18, 2019, we then ran this collection process continuously from December 20, 2019 to March 24, 2020.  Over this entire time period, we were able to collect 
67.6 million tweets, posted by 3.6 million unique Twitter users.

\begin{table}[h!]
	\begin{center}
	    \caption{ \bf Tweet collection terms.} 
        \label{table2}
		\begin{tabular}{|l|l|} 
			\hline
			Keyword   & Date added\\
			\hline
            \#FactsMatter &	2019-12-12 \\\hline
            \#IGHearing	& 2019-12-12 \\\hline
            \#IGReport	& 	2019-12-12 \\\hline
            \#ImpeachAndConvict	& 	2019-12-12 \\\hline
            \#ImpeachAndConvictTrump	& 	2019-12-12 \\\hline
            \#SenateHearing	& 	2019-12-12 \\\hline
            \#TrumpImpeachment	& 	2019-12-12 \\\hline
            impeach		& 2019-12-12 \\\hline
            impeached	& 	2019-12-12 \\\hline
            impeachment	& 	2019-12-12 \\\hline
            Trump to Pelosi		& 2019-12-12 \\\hline
            \#25thAmendmentNow 	& 	2019-12-18 \\\hline
            \#ImpeachAndRemove		& 2019-12-18 \\\hline
            \#ImpeachmentEve		& 2019-12-18 \\\hline
            \#ImpeachmentRally	& 	2019-12-18 \\\hline
            \#NotAboveTheLaw		& 2019-12-18 \\\hline
            \#trumpletter	& 	2019-12-18 \\\hline 
            \#GOPCoverup		& 2020-01-22 \\\hline
            \#ShamTrial		& 2020-01-22 \\\hline
            \#AquittedForever	& 	2020-02-06 \\\hline
            \#CountryOverParty	& 	2020-02-06 \\\hline
            \#CoverUpGOP		& 2020-02-06 \\\hline
            \#MitchMcCoverup		& 2020-02-06 \\\hline
            \#MoscowMitch	& 	2020-02-06 \\\hline
			
		\end{tabular}
	\end{center}
	\begin{flushleft} Keywords used to collect impeachment related tweets, and the date each was added to the collection.
\end{flushleft}
\end{table}

As a part of this tweet collection process, we also collected the Twitter profile of each user. The profile included information such as the name of the user, location (if provided), and a short description provided by the user.


We later used the Twitter Search API to collect user follower networks in a separate collection process.
Our network convention was to have follower edges point from a user to the person that followed them.  This way the follower edges point in the direction of information flow, as tweets from a user appear in the timeline of their followers.
To build the follower network for the users in our dataset, we used a customized web crawler to collect a list of \textit{followings} for each user (i.e. the users they follow). We chose to collect the followings rather than the followers for each user because it reduced our data collection burden.  We observed that the follower count can be much larger than the following count for a Twitter user, especially for the more popular users.  Therefore, to more easily collect all edges in the follower network, we collected the users followings.  To be able to collect the follower network in reasonable time, we collected a maximum of 2,000 followings per user.  {This value was sufficient for our data collection purposes as 85\% of the following counts were below this value.}  In total we obtained {53.4 million edges} in this follower network.

The data and code needed to reproduce our results can be found in the GitHub repositories located at \url{https://github.com/s2t2/tweet-data-2020} and \url{https://github.com/s2t2/tweet-analysis-2020}, respectively.  We have included in the data repository the tweet identifiers if one wishes to recollect the data from Twitter. All data was collected and is shared in accordance with Twitter's Terms of Service.

\subsection*{Partisanship classification model}

The partisanship classifier we used is a bidirectional encoder representations from transformers (BERT) language representation model \cite{devlin2018bert}.  Transformers are a neural network architecture that have shown incredible success in language modeling \cite{vaswani2017attention}.  BERT is a bidirectional transformer pre-trained using a combination of a masked language modeling objective and a next sentence prediction objective on the Toronto Book Corpus (800 million words) \cite{zhu2015aligning} and Wikipedia (2.5 million words).  The BERT model provides a sentence embedding that can be used for many natural language processing tasks such as sentiment classification.  
The tweet text is fed into the BERT model and mapped into an embedding representation.  We use the base BERT model which produces a 768 dimensional embedding.  This representation is then fed to a fully connected single layer of 768 neurons with linear activation.  The linear layer has two outputs which correspond to pro-Trump and anti-Trump sentiment. Our architecture follows the standard use of BERT for sentiment classification. Details on the architecture can be found in \cite{devlin2018bert}.  To obtain the sentiment of the tweet we use the value from the pro-Trump output so strong anti-Trump and pro-Trump sentiment are equal to zero and one, respectively.


We created training data for the model using strongly partisan users.  These users were identified by the content of their Twitter profile descriptions.  If a user's description contained any of the words in Table~\ref{table3} and none of the keywords in Table~\ref{table4}, the user was given a label of zero, indicating anti-Trump sentiment.  The opposite was done to identify pro-Trump users, who were given a label of one.  The labels of these strongly partisan users were assigned to their impeachment related tweets in our dataset.  This process created a labeled training set of over 14 million tweets.

\begin{table}[h!]
	\begin{center}
	    \caption{ \bf Keywords used to identify anti-Trump users. } 
		\label{table3}
		\begin{tabular}{|l|l|}
		\hline
		\centering\textbf{Hashtag} & \textbf{Description}\\\hline
            \#BIDEN2020 & \\\hline
            \#BLM & ``Black Lives Matter'' - a movement for racial equality  \\\hline
            \#BLUEWAVE &   \\\hline
            \#BLUEWAVE2020 &   \\\hline
            \#DEMCAST & A left-leaning media outlet  \\\hline
            \#FBR &  ``Follow Black Resistance'' \\\hline
            \#IMPEACH &   \\\hline
            \#IMPEACHANDREMOVE &   \\\hline
            \#IMPEACHTRUMP &   \\\hline
            \#IMPEACHTRUMPNOW &   \\\hline
            \#IMPOTUS & ``Impeached POTUS''  \\\hline
            \#METOO & A movement for gender equality  \\\hline
            \#NOTMYPRESIDENT	 &   \\\hline
            \#RESIST	 &   \\\hline
            \#RESISTANCE	 &   \\\hline
            \#RESISTER &   \\\hline
            \#THERESISTANCE	 &   \\\hline
            \#VOTEBLUE &   \\\hline
            \#VOTEBLUE2020	 &   \\\hline
            \#VOTEBLUENOMATTERWHO &   \\\hline
            \#WTP2020 & 	``We The People 2020''  \\\hline
	    \end{tabular}
	\end{center}
\end{table}

\begin{table}[h!]
	\begin{center}
		\caption{ \bf Keywords used to identify pro-Trump users.}
		\label{table4}
		\begin{tabular}{|l|l|}%
		\hline
		\centering\textbf{Hashtag} & \textbf{Description}\\\hline
            \#1A & The First Amendment\\\hline 
            \#2A & The Second Amendment\\\hline 
            \#AMERICAFIRST & A Trump campaign slogan\\\hline 
            \#BUILDKATESWALL & \\\hline 
            \#BUILDTHEWALL & A Trump campaign slogan\\\hline
            \#CODEOFVETS & \\\hline 
            \#CONSERVATIVE & \\\hline 
            \#DEPLORABLE & Refers to a Hillary Clinton quote from the 2016 election\\\hline 
            \#DRAINTHESWAMP & A Trump campaign slogan\\\hline 
            \#KAG & ``Keep America Great'' - a Trump campaign slogan\\\hline 
            \#MAGA & ``Make America Great Again'' - a Trump campaign slogan\\\hline 
            \#NRA & The National Rifle Association\\\hline 
            \#PATRIOT & \\\hline 
            \#POTUS45 & 	45th President (Trump)\\\hline 
            \#QANON & 	Related to Qanon conspiracy theory\\\hline 
            \#THEGREATAWAKENING & 	Related to Qanon conspiracy theory\\\hline 
            \#TRUMP & \\\hline 
            \#TRUMP2020 & \\\hline 
            \#TRUMPTRAIN & 	\\\hline 
            \#VETERAN & 	\\\hline  
            \#WALKAWAY & 	\\\hline  
            \#WWG1WGA & Related to Qanon conspiracy theory\\\hline  
	    \end{tabular}
	\end{center}
\end{table}

We used 637,672 of the labeled tweets to train the BERT sentiment classifier   The tweets were chosen so that 50\% were anti-Trump and 50\% were pro-Trump, creating a balanced set of labels.  We did not use all of the labeled tweets because the size of this dataset made the training process very slow.  We found that using a smaller number of tweets resulted in a much faster training process while still producing a highly accurate classifier.    
To prevent over-fitting during training we use a dropout layer on the BERT output with a dropout probability of 0.3.  The classifier was trained for ten epochs over the data using the Adam optimizer \cite{kingma2014adam} and a cross-entropy loss function.  The data was split into 80\% for training, 10\% for validation, and 10\% for testing in a stratified manner.  

\begin{figure}
    \centering
    \includegraphics[width = \textwidth]{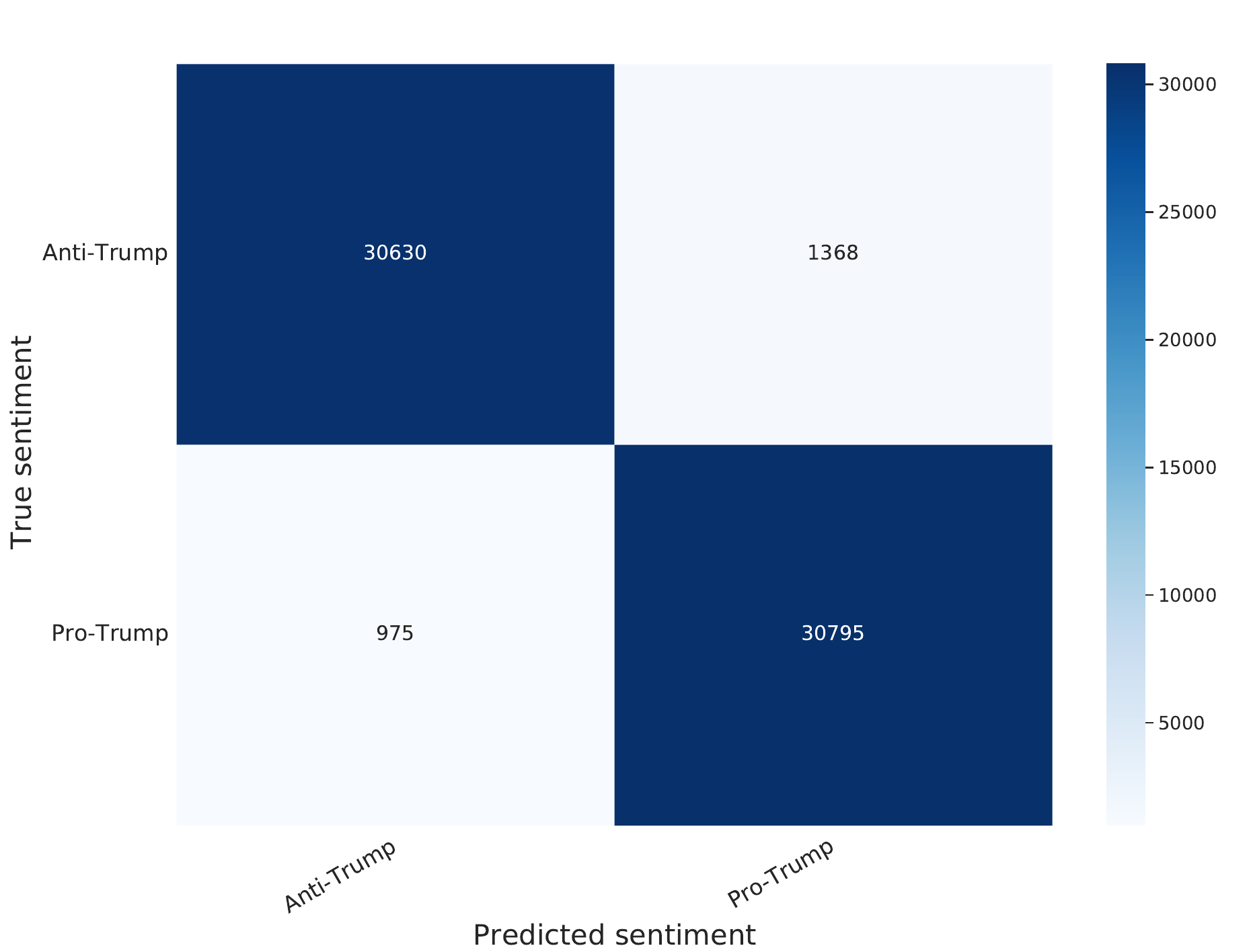}
    \caption{\textbf{Confusion matrix of trained partisanship classifier applied to test data.} }
    \label{fig:confusion}
\end{figure}

\begin{figure}
    \centering
    \includegraphics[width = \textwidth]{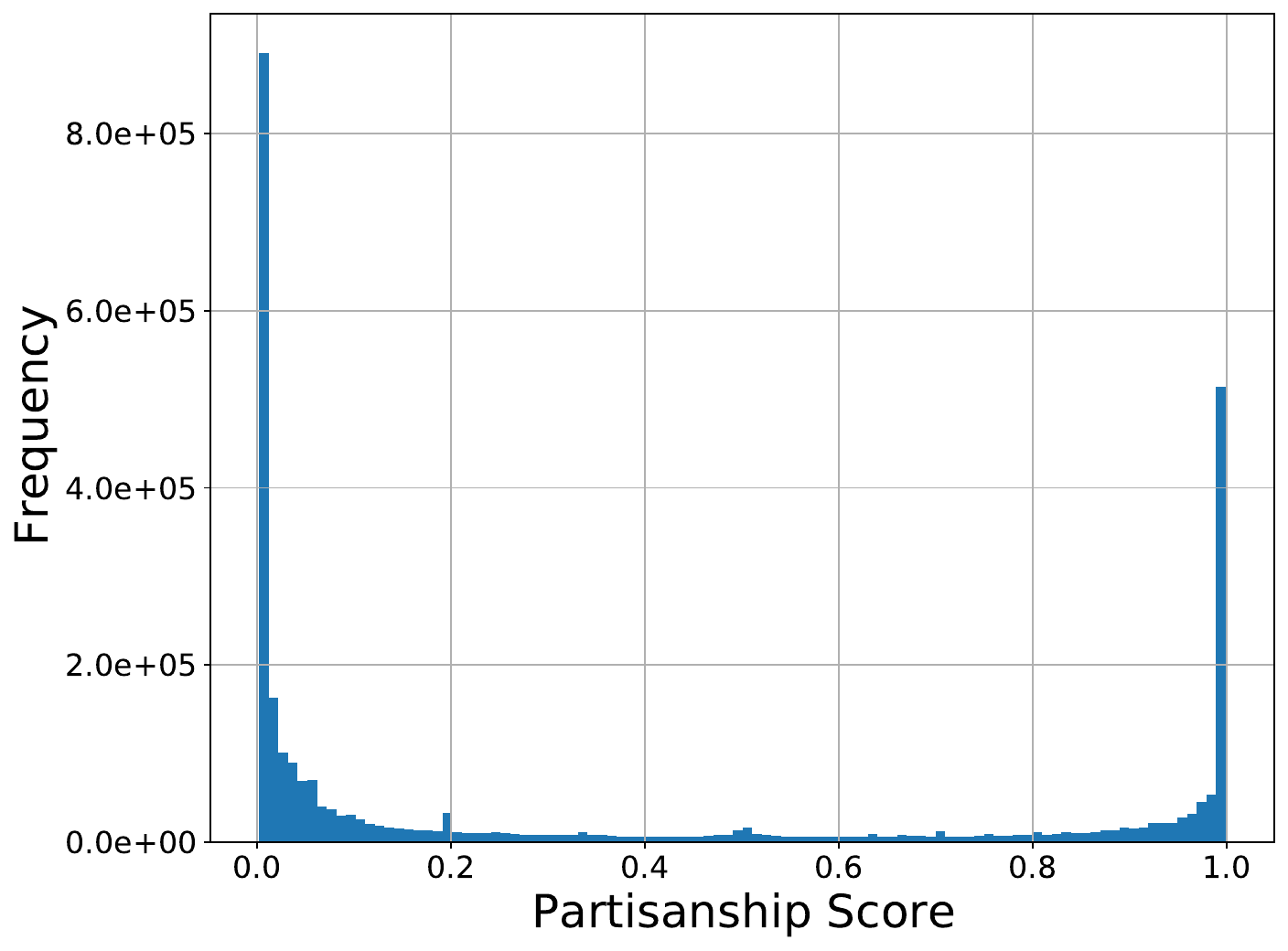}
    \caption{\textbf{Histogram of the partisanship scores of all users.} }
    \label{fig:partisanship_hist}
\end{figure}

The trained classifier is quite effective at measuring opinions about the impeachment.  On the held out testing data it achieved a 96.3\% accuracy score. We show the confusion matrix on the test data in Fig~\ref{fig:confusion}.  As can be seen, the classifier achieves high true positive and negative rates while maintaining low false positive and negative rates.  Also, the rates are nearly equal for both classes.  We provide some random samples of  tweets from each end of the sentiment spectrum and their partisan sentiment measured by the classifier in Table~\ref{table5}. We plot in Fig~\ref{fig:partisanship_hist} a histogram of the partisanship scores of all users in our dataset as measured by the classifier.  The user scores are obtained by averaging the partisanship score of each user's tweets.

\begin{table}[h!]
	\begin{center}
		\caption{ \bf Example partisanship classification scores.}
		\label{table5}
		\begin{tabular}{|p{0.8\textwidth}|p{0.1\textwidth}|}
		\hline
		\centering\textbf{Tweet text} & \textbf{Opinion}\\\hline
		\textit{Mark my words... Trump is starting a war to distract from the Impeachment.} & 0.105
		\\\hline
		\textit{LA Times joins growing list of papers calling for Trump's impeachment} & 0.17
		\\\hline
		\textit{Flynn sentencing - Jan 28  State of the Union -Feb 4  Stone sentencing - Feb 6  Impeachment - Forever} & 0.26
		\\\hline
		\textit{Americans have officially lost any belief in the Democrats’ partisan impeachment sham.} & 0.96
		\\\hline
		 \textit{It’s time for the Senate to end this partisan,impeachment sham once and for all.} & 0.97
		 \\\hline
		 \textit{Pelosi and Dems blew it! Impeachment Sham is Backfiring!} & 0.98
		 \\\hline
	\end{tabular}
	\end{center}
    \begin{flushleft} Tweets from testing dataset and their opinion scores assigned by the BERT sentiment classifier.  The tweets are ordered by opinion score.  An opinion of zero is anti-Trump and an opinion of one is pro-Trump.
    \end{flushleft}
\end{table}

\subsection*{Qanon classification}

To identify Qanon supporters, we utilized a list of terms commonly used by Qanon supporters at the time, which are shown in Table~\ref{table2_new}. The terms on this list are consistent with those discussed by other work in this field \cite{oconnor2020qanon, xu2022network}.
Other researchers have classified Qanon supporters based on hashtags used in their tweets or retweets \cite{xu2022network}, however our method is based on hashtags used in profile descriptions.
Any user account that included at least one of these terms in their profile description  was labeled as Qanon (after excluding any anti-Trump accounts or accounts which tweeted using anti-Trump hashtags). 
A limitation of this approach is that our list of hashtags is likely not comprehensive of all Qanon hashtags, especially as they may evolve over time.

\begin{table}[h!]
	\begin{center}
	    \caption{ \bf Q-anon user profile terms.} 
        \label{table2_new}
		\begin{tabular}{|l|l|} 
			\hline
			Keyword   \\
			\hline
            \#QANON  \\\hline
            \#WWG1WGA  \\\hline 
            \#GREATAWAKENING  \\\hline 
            \#WAKEUPAMERICA  \\\hline 
            \#WEARETHENEWSNOW  \\\hline 
		\end{tabular}
	\end{center}
	\begin{flushleft} Keywords used to identify Qanon supporters.
\end{flushleft}
\end{table}

\subsection*{Bot detection}
Bots exhibit certain traits and behaviors that allow them to be identified.  Some of these
include excessive retweeting and never posting original tweets.  
These behaviors are likely due to the bots' automated nature.
Many algorithms have been developed
for bot detection in Twitter and each has its own strengths.  The algorithm of \cite{des2021detecting} uses a factor graph model that allows for the simultaneous detection of multiple bots based on the collective retweeting behavior of users discussing a specific topic.  The data required by this algorithm is the set of tweets, or more precisely, the set of retweets, about the topic. Another popular algorithm is the machine learning based Botometer (formerly BotOrNot)\cite{davis2016botornot} which utilizes a large amount of data about an individual account, including followers, friends, tweets, and profile, in order to determine if it is a bot.  Botometer is a good algorithm to use when one wants to determine if an individual account is a bot.  However, when one wants to find bots among a large set of users discussing a topic, as is the case in our impeachment analysis, the factor graph algorithm is more convenient.  We can identify bots with this algorithm without having to collect any additional data, which is important given how large our daily active user sets are.  In addition, it has been found that the factor graph algorithm has slightly better performance than Botometer \cite{des2021detecting}. 
 
 The factor graph algorithm first constructs a retweet network based on the retweets in a given set of tweets.  In this retweet network the nodes are the users who tweet and an edge $(u,v)$ pointing from user $u$ to user $v$ means that $v$ retweets $u$.  The edge $(u,v)$ is given a weight $w_{uv}$ which equals the number of times $v$ retweets $u$.  Like with the follower network, the edge direction in the retweet network indicates the flow of information.   The algorithm uses the retweet network structure to calculate a bot probability for each user.   This probability is based on the empirical observations that bots are likely to retweet humans, but unlikely to retweet bots, and humans are likely to retweet humans and less likely to retweet bots \cite{des2021detecting}.  This homophily for the humans and heterophily for the bots is utilized by the algorithm to determine bot probabilities from the structure of the retweet network.
 
\begin{figure}[!h]
\includegraphics[width = \textwidth]{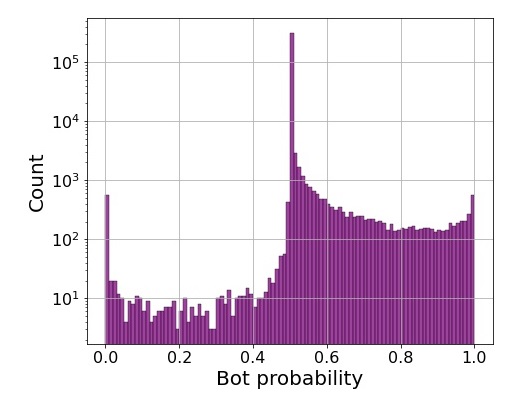}
\caption{{\bf Bot probability classifications on an example day.}
Histogram of the bot probabilities calculated by the factor graph algorithm \cite{des2021detecting} based on the impeachment retweet network on February 1, 2020 (an example day).}
\label{fig8}
\end{figure}

We used the factor graph algorithm to identify bots active each day. To apply the algorithm, we first constructed a retweet network from the tweets posted on a given day.   Once the retweet network is constructed, we apply the algorithm to the network using the parameters specified in \cite{des2021detecting} to simultaneously obtain the joint bot probability for all users. We show one example of the bot probability distribution for a single day in Fig~\ref{fig8}. As can be seen, the bulk of users have bot probabilities near 0.5, which is the algorithm saying it cannot determine one way or another what the user is. In the upper probability range we see that the histogram decreases up to approximately 0.8, and then increases afterwards.  This suggests there is a cluster of nodes with bot probabilities in the interval $[0.8,1.0]$.  We used the lower probability bound of this cluster as the bot probability threshold so that  this cluster of nodes is identified as bots.
  
We found  a similar behavior in the bot probability distribution across all days.  Therefore, we used the same bot probability threshold of 0.8 for each day.  Any user who had a bot probability over the threshold in at least one day in our dataset was declared to be a bot. We use this approach because due to automation, it is easy for a bot to act like a human (retweet less frequently).  However, it is harder for a human to behave like a bot (retweet at an extremely high rate).  Also, bots may intentionally behave in this non-constant manner.  A bot may be extremely active on certain days when it is amplifying certain users, and quiet or only slightly active on others.  Humans showing elevated retweet rates may have difficulty being as active as a bot.  Under these assumptions, exhibiting bot behavior on at least one day would indicate the account is a bot, and humans would not show bot-like behavior any day. Across all days we found {24,150 bots} among {3.6 million} active accounts. As a terminology note, in this paper when we refer to the word ``bot'' we mean the accounts likely to be retweet bots, and when we refer to the word ``human'' we mean accounts not likely to be retweet bots (i.e. non-bots).

\begin{figure}[!h]
\includegraphics[width=\textwidth]{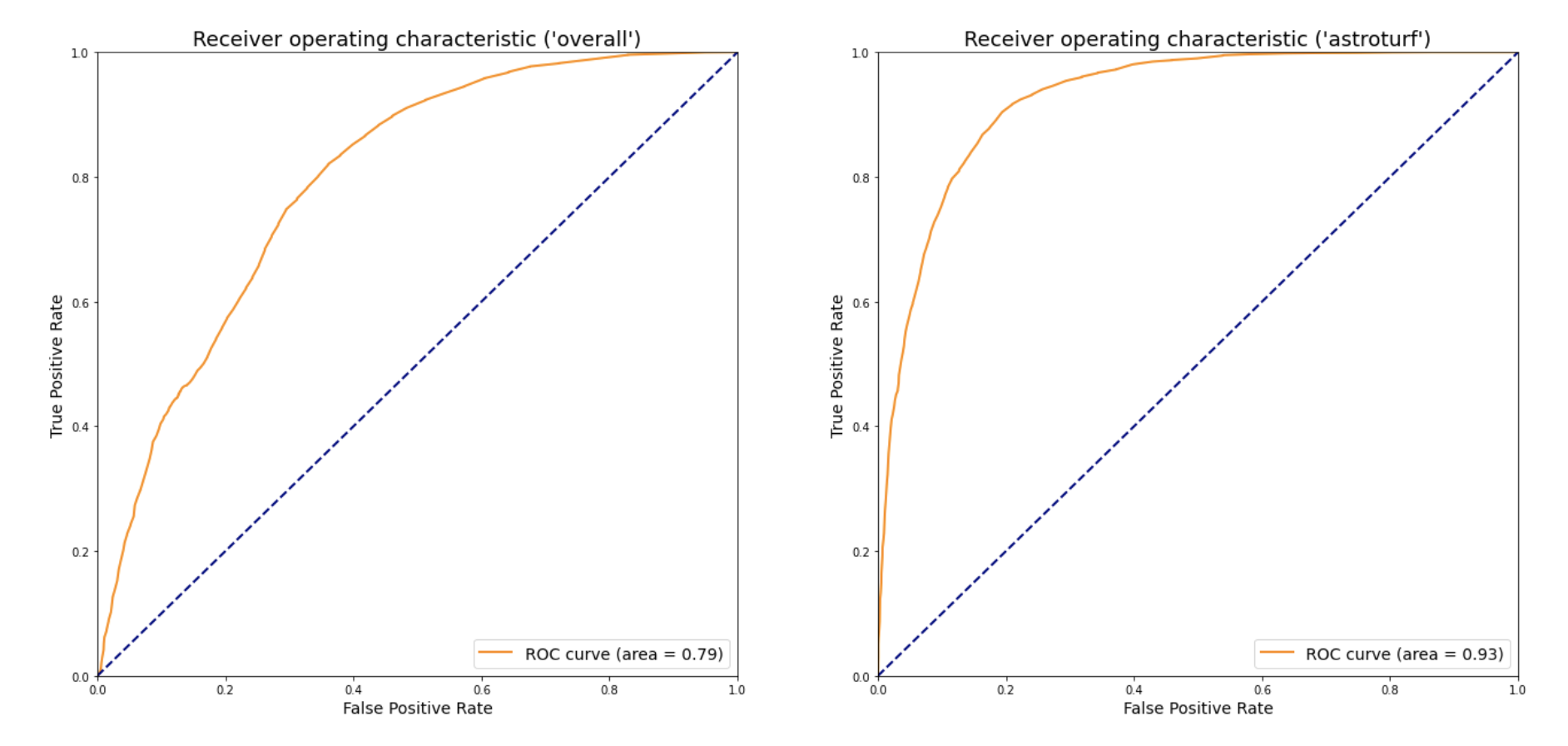}
\caption{{\bf ROC curves for (left) ``overall'' Botometer scores and (right) ``astroturf'' Botometer scores  using our bot classification results as ground truth.}}
\label{fig9}
\end{figure}

To explore the validity of our bot classifications, we took a random sample of around 7500 accounts with a roughly 60/40 human to bot ratio, and compared them against bot scores obtained from the Botometer API \cite{sayyadiharikandeh2020detection}. The Botometer API provides scores from 0 to 1 representing the likelihood that a given account is a bot (i.e. ``overall'' and ``cap'' scores), as well as separate scores for a number of bot sub-types (e.g. ``astroturf'', ``financial'', etc.). We construct receiver operating characteristic (ROC) curves for different Botometer scores and our labels as a ground-truth.  We use area under the curve (AUC) score to measure the agreement between our bot classifications and each kind of score provided by Botometer.  We find that the AUC score for the ``overall'' bot scores is around 0.79, while the AUC score for the ``astroturf'' scores is much higher, around 0.93 (see Fig~\ref{fig9}).  This suggests that our bot detection algorithm, when used within the context of a political discussion, performs generally well at identifying bots overall, and performs even better at identifying hyper active political bots.

\subsection*{Distribution of toxicity scores}
\begin{figure}[!h]
\includegraphics[width = \textwidth]{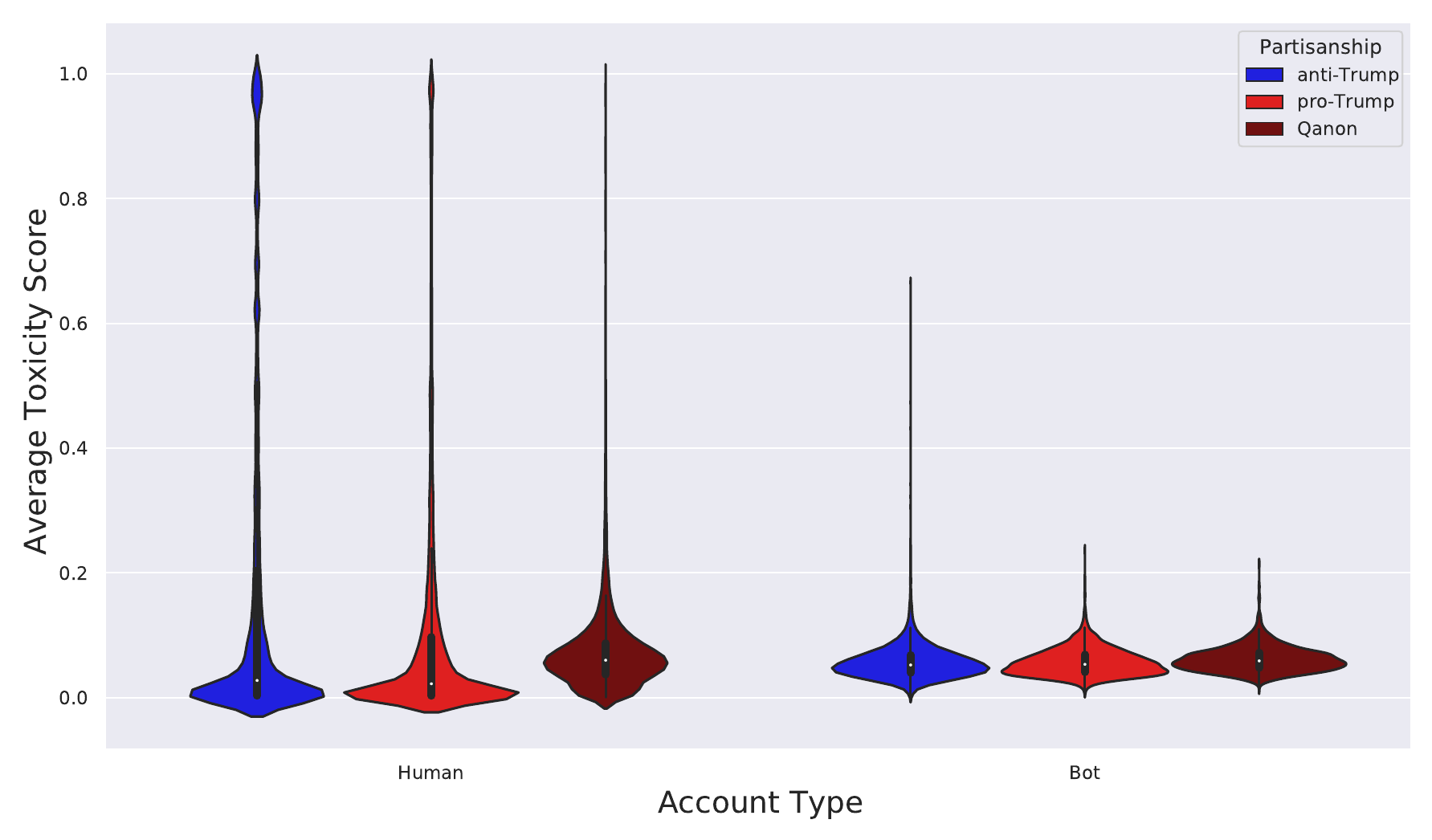}
\caption{{\bf Violin plot of the mean user toxicity score for different user types.}}
\label{fig10}
\end{figure}

Fig~\ref{fig10} shows a violin plot of the mean user toxicity score for the different user types.  From this figure we see that there are clusters of humans with very high toxicity scores.  The bots do not have such a toxic cluster. To further quantify these clusters, let us define a toxic user as one whose average toxicity score is greater than or equal to 0.9.  With this definition, we find that 2\% of Republican humans are toxic, while 6\% of Democrat humans are toxic.  There are zero toxic bots under this definition. 

\subsection*{Generalized harmonic influence centrality calculation}
To calculate the generalized harmonic influence centrality (GHIC) of a set of nodes we need several pieces of information.  First, we need the follower network of the nodes.  We use the daily active follower network for this purpose, which is the follower network induced by users who tweet about the impeachment on a given day.  

Second, we need to characterize the activity level of the users.  We define this to be the tweet rate of the users.  We measure the tweet rate as the total number of impeachment tweets posted divided by the duration of our data collection.  This gives a stable measure of the posting rate of a user.  We use this same posting rate for each daily GHIC calculation as it provides an accurate measure of the general activity level of a user.

Third, we need to know the political sentiment of each user.  We obtain this by taking the average sentiment of a user's tweets as measured by our BERT partisanship classifier.

Fourth, we need to identify a subset of users in the daily active follower network as stubborn.  GHIC assumes that the stubborn users are not persuadable and their sentiment does not change.  We first set all bots to be stubborn, as they are automated accounts that do not respond to persuasion.  We identify stubborn users among the humans based on their political sentiment or opinion.   Studies have shown that stubborn users have extreme opinions \cite{moussaid2013social}.  To operationalize this principle, we follow the approach used in \cite{des2021detecting, hunter2018opinion} to define extreme opinions as those below the 10th percentile and above the 90th percentile of the opinions of all users in our dataset. It has been shown that the GHIC is robust to the precise value of these thresholds \cite{des2021detecting}.

Once the requisite information has been obtained, the GHIC of a set of nodes can be calculated. We present the steps for GHIC calculation here, drawing from the presentation in \cite{des2021detecting}. We are given a follower network $G = (V,E)$ with node set $V$ (the Twitter users) and edge set $E$ (the follower edges).  We assume each node follows a set of users that we define as their \emph{following}.  Each node $v\in V$ posts content (tweets) at a rate $\lambda_v$.  We define the stubborn users as the set $V_0\subset V$ and the non-stubborn users as $V_1\subset V$.  We define $\mathbf \Psi$ as the vector of stubborn opinions and $\mathbf \theta$ as the vector of non-stubborn opinions.  We are given the stubborn opinions $\Psi$ and we obtain the non-stubborn opinions $\theta$ by solving  
\begin{equation}
\mathbf G \mathbf\theta = \mathbf F\mathbf \Psi,\label{eq:equilibrium_matrix}
\end{equation}
where the matrix $\mathbf G$ is given by the equation:
\begin{align*}
\mathbf G_{ij} = \begin{cases}
-\sum_{k\in\text{following of~}i} \lambda_{k}& \quad i=j, i\in V_1 \\
\lambda_{j} & \quad i\neq j, (j,i)\in E, i,j \in V_1 \\
0 & \text{else}, \\
\end{cases}
\end{align*}
and the matrix $\mathbf F$ is given by
\begin{align*}
\mathbf F_{ij} = \begin{cases}
-\lambda_{j} & \quad (j,i)\in E, i\in V_1, j\in V_0  \\
0 & \text{else}. \\
\end{cases}
\end{align*}
Equation \eqref{eq:equilibrium_matrix} is the equilibrium opinions in the opinion dynamics model upon which GHIC is based.  

Next we choose a set of nodes $S\subset V$ for which we want to calculate the GHIC.  We define a new network $G' = (V',E')$ where $V' = V/S$ and $G'$ is the sub-network of $G$ induced by $V'$.  $G'$ is the network $G$ but with the nodes in $S$ removed.  Let $\theta$ and $\theta'$ be the solution of equation \eqref{eq:equilibrium_matrix} assuming the underlying network is $G$ and $G'$, respectively.  The GHIC of $S$ is then given by 

\begin{align*}
\text{GHIC}(S) & = \frac{1}
{\vert V_1/S\vert }\sum_{i\in V_1/S} \theta_i - \theta_i'.
\end{align*}
We see from this that the GHIC of $S$ is the change in mean non-stubborn equilibrium opinion caused by the presence of 
the $S$ nodes in the network.

\subsection*{Robustness of generalized harmonic influence centrality}

\begin{figure}
    \centering
    \includegraphics[width =\textwidth]{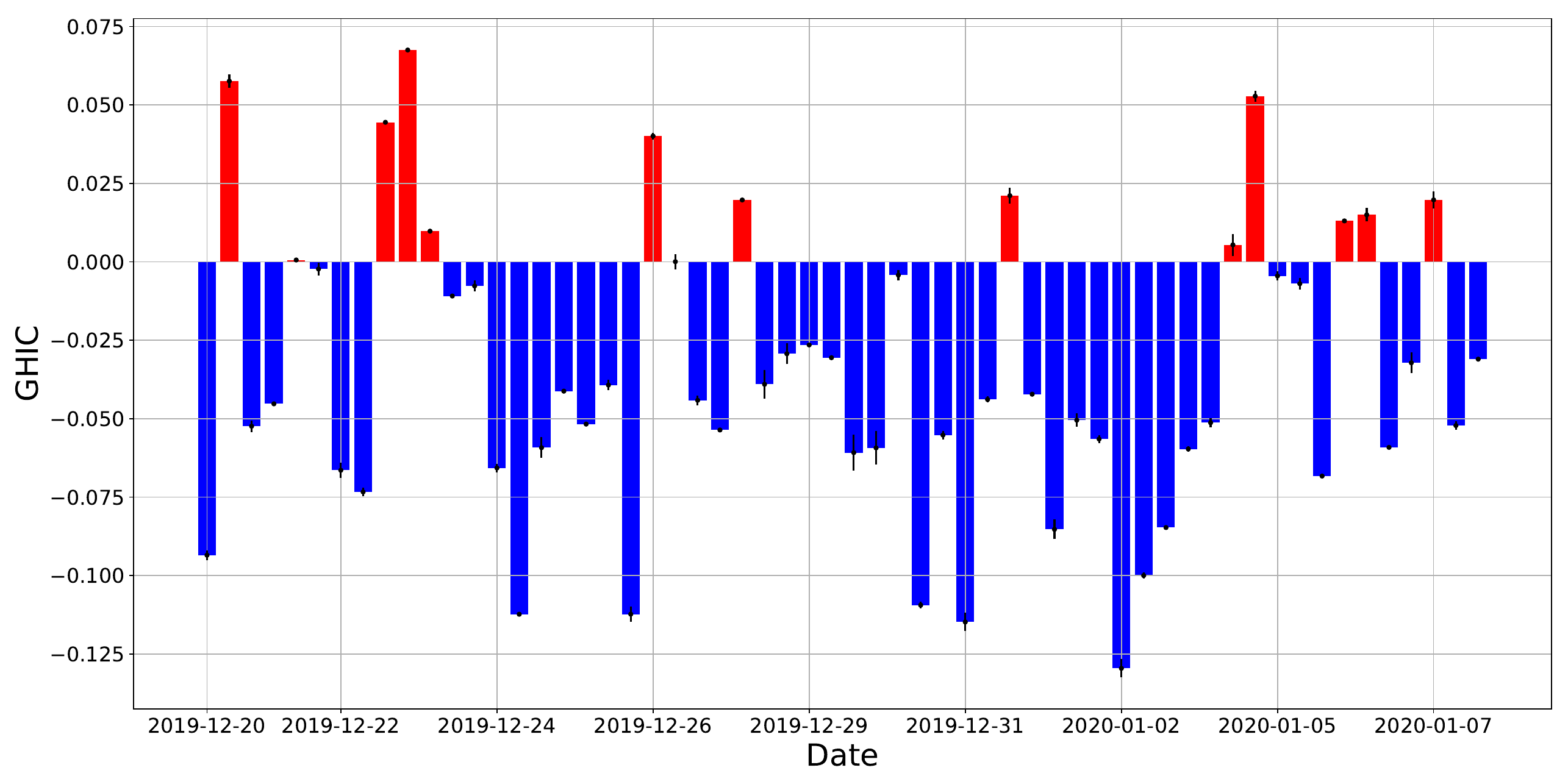}
    \caption{Daily generalized harmonic influence centrality (GHIC) score for all bots versus date. 
 The error bars correspond to the range of GHIC values for bot probability thresholds of 0.72, 0.80, and 0.88.}
    \label{fig:ghic_robust_bot}
\end{figure}

We perform checks to demonstrate the robustness of the  generalized harmonic influence centrality (GHIC) with respect to the bot probability threshold.  Bots are identified as accounts whose bot probability, as determined by our algorithm, exceeds a threshold of 0.8.  We test different values for this threshold and see how the GHIC is affected.  We chose to test the values 0.72 and 0.88, which are 10\% above and below the baseline threshold.  The results of this analysis are shown in Fig~\ref{fig:ghic_robust_bot}.  Changes in the bot probability threshold do not cause substantial changes in the GHIC, as can be seen from the narrow error bars.  We find that the median absolute shift of the GHIC with respect to the bot threshold is 6\%, suggesting that the GHIC is robust to the bot probability threshold.

\nolinenumbers


%
%
%



%
%
 \bibliography{final} 

\end{document}